# Convolutional neural network models for cancer type prediction based on gene expression

Milad Mostavi[1,2], Yu-Chiao Chiu[1], Yufei Huang[2,3,§], Yidong Chen[1,3,§]

[1]Greehey Children's Cancer Research Institute, University of Texas Health Science Center at San Antonio, San Antonio, TX 78229, USA

[2]Department of Electrical and Computer Engineering, University of Texas at San Antonio, San Antonio, TX 78249, USA

[3]Department of Epidemiology and Biostatistics, University of Texas Health Science Center at San Antonio, San Antonio, TX 78229, USA

[§]Corresponding authors

Email addresses:

YH: Yufei.Huang@utsa.edu

YC: ChenY8@uthscsa.edu

# Abstract

**Background**

Precise prediction of cancer types is vital for cancer diagnosis and therapy. Important cancer marker genes can be inferred through predictive model. Several studies have attempted to build machine learning models for this task however none has taken into consideration the effects of tissue of origin that can potentially bias the identification of cancer markers.

**Results**

In this paper, we introduced several Convolutional Neural Network (CNN) models that take unstructured gene expression inputs to classify tumor and non-tumor samples into their designated cancer types or as normal. Based on different designs of gene embeddings and convolution schemes, we implemented three CNN models: 1D-CNN, 2D-Vanilla-CNN, and 2D-Hybrid-CNN. The models were trained and tested on combined 10,340 samples of 33 cancer types and 731 matched normal tissues of The Cancer Genome Atlas (TCGA). Our models achieved excellent prediction accuracies (93.9-95.0%) among 34 classes (33 cancers and normal). Furthermore, we interpreted one of the models, known as 1D-CNN model, with a guided saliency technique and identified a total of 2,090 cancer markers (108 per class). The concordance of differential expression of these markers between the cancer type they represent and others is confirmed. In breast cancer, for instance, our model identified well-known markers, such as *GATA3* and *ESR1*. Finally, we extended the 1D-CNN model for prediction of breast cancer subtypes and achieved an average accuracy of 88.42% among 5 subtypes. The codes can be found at

https://github.com/chenlabgccri/CancerTypePrediction.

## Conclusions

Here we present novel CNN designs for accurate and simultaneous cancer/normal and cancer types prediction based on gene expression profiles, and unique model interpretation scheme to elucidate biologically relevance of cancer marker genes after eliminating the effects of tissue-of-origin. The proposed model had light hyperparameters to be trained and thus can be easily adapt to facilitate cancer diagnosis in the future.

**Keywords:** Deep Learning; Convolutional Neural Networks, The Cancer Genome Atlas; Cancer type prediction; Cancer gene markers; Breast cancer subtype prediction

## Background

Cancer is the second leading cause of death worldwide, an average of one in six deaths is due to cancer [1]. Considerable research efforts have been devoted to cancer diagnosis and treatment techniques to lessen its impact on human health. Cancer prediction's major focus is on cancer susceptibility, recurrence, and prognosis, while the aim of cancer detection is the classification of tumor types and identification of markers for each cancer such that we can build a learning machine to identify specific metastatic tumor type or detect cancer at their earlier stage. With the increased awareness of precision medicine and early detection techniques matured over years of technology development [2-4], including particularly many detection screens achieving a sensitivity around 70-80% [5], the demand for applying novel machine learning methods to discover new biomarkers has become one of the key driving factors in many clinical and translational applications.

Deep learning (DL), a branch of Artificial Intelligence, is a family of multi-layer neural network models that excel at the problem of learning from big data [6]. Similar to other machine learning methods, DL consists of the training step where the estimation of network parameters from a given training dataset is carried out, and the testing step that utilizes the trained network to predict outputs of new input data. The accumulation of whole transcriptomic profiling of tumor samples enabled the pursuit of DL model for better accuracy and innovative interpretability for cancer type prediction. One prominent resource of cancer transcriptomic profiling is The Cancer Genome Atlas (TCGA) which consists of more than 11,000 tumors from 33 most frequent cancer types[7]. Several DL models have been developed for early cancer diagnosis. Ahn, *et al.,* [8] designed a fully connected deep neural network (DNN) trained using a dataset of 6,703 tumor and 6,402 normal samples, and provided an initial assessment of individual gene's contribution to the final classification. Lyu, *et al.,* [9] and Li *et al.* [10] extended such effort to classifying individual tumor types. Li *et al.* proposed a *k*-nearest neighbors (KNN) algorithm coupled with a genetic algorithm for gene selection and achieved >90% accuracy for predicting 31 cancer types. Lyu *et al.* proposed a CNN model with 2D mapping of the gene expression samples as input matrices and achieved >95% accuracy for all 33 TCGA cancer types. Lyu *et al.,* also provided a data interpretation approach based on Guided Grad-Cam [11]. GeneCT [12] is another attempt which constrains the input genes to 2 categories: oncogenes and tumor suppressors (1,076 genes in total) to determine the cancerous status that includes transcription factors (1,546 genes) to classify samples to the tissue of origin. The paper reported an overall accuracy of 97.8% with the 10-fold cross-validation. Instead of using transcriptomic data, DeepCNA [13], a CNN based classifier, utilized ~15,000 samples with copy number aberrations (CNAs) from COMICS [14] and the HiC data from 2 human cell-lines and achieved an accuracy ~60% to discern 25 cancer types. While all these attempts achieved high accuracy to some extent, these methods ignore the existence of tissue of origin within each cancer type. Without removing the influence of normal tissues during cancer classification, the implementation of data interpretation scheme will unlikely to differentiate tissue-specific genes or cancer-type-specific genes. Thus, it is impossible to perform functional analysis or select biomarkers for cancer detection from such models. Morever, none of these studies systematically evaluated different CNN model constructions and their impact on the classification accuracy.

In one of our earlier attempts [15], Chen, *et al*, constructed an autoencoder system (GSAE) with embedded pathways and functional gene-sets at each input node to reduce the number of weights to be estimated. They applied the GSAE to classify breast cancer subtypes. Here we presented a study of different CNN models constructed for different input data formats. These models systematically interrogate the capacity of the convolution kernels. Utilizing the entire collection of TCGA gene expression data sets, covering all 33 cancer types and nearly 700 normal samples from various tissues of origin, we examined the accuracies of tumor type prediction before and after removing the influence of tissue-specific genes' expression. In addition, we proposed a unique model interpretation scheme to examine the impact of all key genes participated in the DL prediction machinery and demonstrated the unique characteristics of the proposed CNN models and the feasibility of extracting diagnostic markers for future validation studies.

## Methods

**Datasets**

We downloaded pan-cancer RNA-Seq data from The Cancer Genome Atlas (TCGA) [16] by an R/Bioconductor package TCGAbiolinks [17] in December 2018. The dataset contained 10340 and 731 samples for 33 cancer types and 23 normal tissues, respectively. We represented gene expression by $log_2(FPKM + 1)$, where $FPKM$ is the number of fragments per kilobase per million mapped reads. Genes with low information burden (mean < 0.5 or st. dev. < 0.8) across all TCGA samples, regardless their cancer types, were removed. We specifically chose a collection of relative higher overall expression and high variable genes in order to reduce the number of non-informative, or noise-sensitive features, within the dataset. Total of 7,091 genes remained after the filtering step. In order to round inputs dimension and facilitate modeling part, nine zeros were added to the gene expressions for having vectors with the length of 7,100. We also collected the PAM50 subtypes of 864 breast cancer (BRCA) samples from TCGA [16]. To test the robustness of our models, we added Gaussian noises with zero mean and standard deviations of 0-500% ($k$) of $i^{th}$ gene average ($\mu_i$), or $N(0, k\mu)$ to each gene. We set noisy gene expression level to 0 if noise added expression level is less than 0.

**Proposed models**

Different CNN models were proposed for cancer type prediction. Each model aims to address a specific aspect of modeling the gene expression data. Few methods were proposed earlier to address input gene order and optimizing the arrangement of genes that leads to best prediction results in [9] where genes were ordered by their chromosomal positions. In this paper, we kept genes in one preset order but instead, exploit the design of CNN kernels to learn correlations among genes. The other consideration is the depth of the CNN. Although deeper CNN models are known to produce more accurate classifications in computer vision [6], several studies have shown that increasing the depth of CNN models on biological data does not always lead to improvement on performance [18]. Here we constrained our designs to include only one layer of convolution. In fact, shallower models are preferred for problems such as cancer type prediction, where there are limited samples relative to the number of parameters. Such shallow models avoid overfitting and also demand fewer resources for training[19, 20]. Based on these two considerations, we discuss three different CNN designed next.

**CNN with vectorized input.** This CNN model takes the gene expression as a vector and applies one-dimensional kernels to the input vector. The output of 1-D convolutional layer is then passed to a maxpooling layer, a Fully Connected (FC) layer, and a prediction layer (**Figure 1A**). For the sake of simplicity, we call this model *1D-CNN*. The main nuance between the proposed 1D-CNN and other counterpart CNNs for applications such as time series prediction is that the stride of the convolution is same as the length of kernel size. As a matter of fact, in some applications, 1D CNN is harnessed to capture temporal relationships between adjacent values in the input. However, in our case, since we are not confident that there are correlations among neighboring gene expression values in the input vector, we choose the stride of CNN as big as the kernel size to capture only the global features associated with this kernel.

**CNN with matrix input.** The second CNN model follows the most commonly practiced types of CNN applications in computer vision where the input has a 2-D format like an image. This CNN includes 2D kernels to extract local features in the input **Figure 1B**. Similar to [9], we reshaped the input gene expression into the 2D space without any specific arrangement to construct an image-like input before feeding it to the 2D CNN. The 2D CNN includes the convolutional layer with the 2D kernel, a maxpooling layer, an FC layer, and a prediction layer. For convenience, we term this model as the *2D-Vanilla-CNN*.

**CNN with matrix input and 1D kernels**. The third model is the *2D-Hybrid-CNN*, which is inspired by the parallel towers in the Resnet modules [21] and simplicity of our proposed 1D-CNN. It is proposed to take advantage of having 2-D inputs with simple 1D convolution operations. In this model, as can be seen in **Figure 1C**, two 1D-kernels slide over the inputs, where one with the size of a row slides vertically and the other one with the size of a column slides horizontally across the 2D input. The outputs of two 1D-kernels are then passed through a maxpooling layer before being concatenated and fed into the FC and prediction layers. As in the Resnet modules, we believe this design can capture more global unstructured features in the input gene expression.

**Implementation of 2D-3Layer-CNN.** In We implemented the model proposed in [9] with all details in Keras DL platform and named it 2D-3Layer-CNN in order to have a fair side-by-side comparison between CNN models developed in this paper. This model contains three CNN modules which in each one Batch Normalization, Activation Function (AF), and Maxpooling are used in a cascade manner. The output of last CNN module is fed into two FC layers and finally softmax layer is used for predicting 33 different cancer types.

**CNN Model Interpretation**

We utilized guided gradient saliency visualization method provided by the Keras visualization package keras-vis [22]. This method calculates the output gradient classes with respect to a small change in gene expressions. The positive values of these changes prime us the importance of those gene expressions in the inputs [23]. In the saliency map generation step, each sample was fed into the model to construct an interpretation map. We then summarized each cancer type as well as for the normal samples by averaging across all samples of the group and constructed a gene-effect matrix of 7,091x34 (33 cancer type and one normal class) that contains gene-effect scores with range of [0, 1] with 1s have maximum effect and 0 to no effect. A gene with a gene-effect score greater than 0.5 was defined as a marker gene for a cancer.

# Results

**Model construction, hyperparameter selection and training**

All of the three models were implemented by Keras [24] DL platform. All of the codes can be found at https://github.com/chenlabgccri/CancerTypePrediction. The input for 1D-CNN (**Figure 1A**) is a 1D vector following gene symbol's alphabetic order, while inputs for 2D-Vanilla-CNN and 2D-Hybrid-CNN (**Figure 1B,C**) models were reshaped to 100 rows by 71 columns matrix. Four of the key hyperparameters known as the number and size of kernels, the stride of kernels, and the number of nodes in the FC layer were tuned by Grid search method provided in [25]. The **Tables 1A** and **B** show all sets of parameters were chosen for 1D-CNN and 2D-Vanilla-CNN models respectively, and their statistical measures on train and test pools. In addition, Categorical Cross Entropy as the loss function, Categorical accuracy as training metric and the Adam optimizer were selected for all 3 CNN models. The epoch and batch size were chosen as 50 and 128, respectively, with the early stopping set with patience = 4 to stop the learning in the case that categorical accuracy did not improve in four consecutive epochs. Finally, ReLU was used as the AF and softmax as the prediction layer at the final layer for of all the models.

All three CNN models were trained with all 10,340 tumor samples initially. To evaluate the training procedure and their robustness against overfitting, we examined loss functions for 3 models **Figure 2A** using 80%-20% splitting for training and validation, and we observed converges to ~0 loss after 10 epochs (where validation's loss at about 0.10 with no obvious overfitting). The model in [9] was trained and tested with same procedure. As can be seen in the **Figure 2A**, the convergence of this model is slower than all proposed three models in this paper.

In order to avoid the bias impacted by stochastic dependency nature of neural networks during training, the 5-fold cross validation was repeated six times (due to the time constraint) and their mean and standard deviation of the classification accuracy were reported for all models. **Figure 2B** (light blue bars) showed classification accuracy at 95.5±0.1%, 94.87±0.04%, 95.7±0.1% for 1D-CNN, 2D-Vanilla-CNN and 2D-Hybrid-CNN, respectively.

**Assessing the impact of tissue-specific features on cancer type prediction**

Considering the tissues of origin when classify tumor samples, previous studies either omitting this important factor by only training the DL machine with tumor samples and then looking for cancer driver genes [9], or training two models: with only cancer associated genes (tumor DL model) or transcription factors (normal DL model) [10]. To observe the influence of tissues of origin with DL model trained with tumor sample only, we fed all 731 normal samples that matched to 23 TCGA cancer types into 1D-CNN model trained on 33 cancer types in the previous section. As is shown in **Figure 2C,** 19 of 23 normal classes are classified into their corresponding cancer type, where normal samples from kidney (KICH, KIRC and KIRP), liver (CHOL and LIHC), lung (LUAD and LUSC) or digestive system (ESCA and STAD) are clearly grouped together, indicating a strong possibility that DL machine was partially trained to recognize tissues of origin. When we examined the classification results of tumor samples (**Figure 2D**), the major classification errors are also within kidney, lung (both boxed in Figure 2D), colon and rectum adenocarcinomas.

**Predicting cancer types without the influence of tissue of origin.**

In order to take into account the impact of tissue of origin in the model, we introduce a new label in the prediction layer where it takes all normal samples (regardless of their original tissue type designation). The $34^{th}$ node in the prediction layer is responsible for removing the trace of tissue of origins from cancer samples, with the intention of achieving a robust cancer type prediction. All three models were re-trained with 33 nodes for tumor classes plus one node for normal samples (labeled as "Normal") with the same architectures correspondingly. Similar to model training with 33 cancer-types only, we had a consistent learning curve (**Figure 3A**) using 80%-20% splitting for training and validation, and converged to ~0 loss after 10 epochs without obvious overfitting. As shown in **Figure 2B** (brown bars), we achieved the overall accuracies 94.9±0.1%, 93.9±0.6%, 95.0±0.1% for 1D-CNN, 2D-Vanilla-CNN and 2D-Hybrid-CNN, respectively, slightly lower than 33 cancer only training, due to the introduction of normal samples (Precision at 92.5%, **Figure 3B**).

Further evaluation of micro-averaged precision-recall statistics of 1D-CNN model with 34 output nodes yielded some interesting observations (**Figure 3B**). The DL machine has a large discrepancy in precision-recall value of tumor type READ. This is due to the large number READ (rectum adenocarcinoma, 83) samples misclassified into COAD (colon adenocarcinoma), causing much lower recall level (48.8%) (**Figure 3C**), while 37 COAD samples are misclassified into READ types. Cholangiocarcinoma (CHOL) has only 36 tumor samples total but a large fraction misclassified into hepatocellular carcinoma (LIHC, 3 samples (~9%)) and Pancreatic Adenocarcinoma (PAAD, 2 samples). Cholangiocarcinoma is a bile duct cancer, and specifically the distal region (extrahepatic cholangiocarcinoma) is made up of the common bile duct that passes through the pancreas, thus potentially the cause of misclassification. We have attempted to train with a separated kidney normal tissue group with no clear improvement (data not show). Evidently, the limited normal samples per tumor group leaves a lot of room to improve.

**Interpretation of the 1D-CNN model to investigate cancer marker genes**

We systematically investigated the 1D-CNN model to understand how the model predicted cancer types with the aim to identify cancer marker genes. The interpretation was accomplished by generating the saliency map (see Methods Section) of 1D-CNN model.

**Interpretation of the 1D-CNN model to investigate cancer marker genes**. We systematically interpreted the 1D-CNN model to get insight on how the model predicts cancer types and finally identify cancer marker genes. We first examined the distribution of gene-effect scores of saliency maps for all caner types, and generally they followed the power law (**Figure 4A**). We set criteria on the gene-effect scores to identify marker genes (see Methods). t-SNE plots on expression data of selected marker genes confirmed that the identified markers preserved the differences among classes even when stringent thresholds were set (score > 0.5 to > 0.9 yielding 2,090 to 91 unique marker genes, respectively; (**Figure 4B**). To include more potential cancer markers into the investigation, we used the threshold of 0.5 for subsequent analyses. We obtained a total of 3,683 markers (2,090 unique genes) for all the 34 classes with minimum of 4 markers to maximum of 346 (**Figure 4C**), or average ~108 markers per cancer type. Diffuse large B-cell lymphoma (DLBC), breast invasive carcinoma (BRCA), and prostate adenocarcinoma (PRAD) were found with the most markers (346, 323, and 230, respectively). Interestingly, the cancers that our model tended to confuse, such as lung cancers (adenocarcinoma [LUAD] and squamous cell carcinoma [LUSC]) and rectum adenocarcinoma (READ), had a much smaller number of markers (4, 4, and 8, respectively). The finding suggested our model's low confidence in classifying cancer types with few marker genes and the requirement of additional modes of genomics profiles (methylation, etc.) to further discriminate cancer types within the same tissue of origin.

**Discrimination capability of marker genes.** We investigated whether simple linear-like differential expression between classes underlying the capacity of these marker genes. The 99 marker genes with a gene-effect score > 0.5 obtained from the normal class indeed had significantly larger differences in the expressional level between pan-cancers and normal samples than other genes ($t$-test $P = 1.4 \times 10^{-3}$; **Figure 4D**), though the differences were moderate in magnitude (mean, 0.55 vs. 0.43). Taking BRCA as a demonstrating example, 323 BRCA markers had a larger differential expression between BRCA and other cancer samples than 6,768 non-marker genes ($P = 3.6 \times 10^{-8}$; **Figure 4E**). The phenomenon held for all markers of any cancer types ($P = 1.6 \times 10^{-47}$). Taken together, our model indeed identified genes with differential expression between classes.

**Marker genes in the breast cancer group.** We further examined a well-studied cancer type, BRCA, as a demonstrating example to the marker genes identified by our model. BRCA had 323 marker genes (gene-effect score > 0.5). Well-known specific markers of BRCA, such as *GATA3* [26] and *ESR1* [27] were ranked at the 13[th] and 98[th] among all genes. Their classifying capability was predominately in BRCA (gene-effect scores, 0.89 and 0.67; **Figure 4F**). Also, we identified other promising novel markers of BRCA, such as *GPRIN1* (the top marker gene with a score of 0.97; **Figure 4F**), *EFNB1* (2[nd], score = 0.94), and *FABP4* (3[rd], score = 0.92), that warrant further investigations.

**Biological functions of marker genes.** To understand biological functions underlying cancer classification, we performed a functional annotation analysis on marker genes of each cancer type or normal. Each set of marker genes were systematically tested for enrichment in a chemical and genetic perturbation signature (the CGP collection) curated by the Molecular Signature Database (MSigDB)[28, 29]. With a criterion on one-tailed Fisher's exact test at $P < 0.001$, we identified a total of 32 associated functions among the 34 classes (**Figure 4G**). Among the top function-class pairs we identified several known cancer functions. For instance, a signature identified from a soft tissue cancer, 'NIELSEN SCHWANNOMA DN'[30], was significantly associated with markers of sarcoma (SARC) (top 2$^{nd}$ significant function-class pair; $P = 3.3 \times 10^{-5}$). Also, marker genes of prostate adenocarcinoma (PRAD) were associated with a signature of androgen response, 'NELSON RESPONSE TO ANDROGEN UP' [31]($P = 5.8 \times 10^{-4}$). We also identified several novel marker functions of cancers, such as 'BASSO CD40 SIGNALING UP' in testicular germ cell tumor (TGCT) (1$^{st}$; $P = 2.0 \times 10^{-5}$), and ''WAKABAYASHI ADIPOGENESIS PPARG BOUND 8D' in bladder urothelial carcinoma (BLCA) (3$^{rd}$, $P = 4.1 \times 10^{-5}$). Overall, functional annotation analysis validated what we expect and potentially revealed several novel mechanisms through the CNN model interpretation. However, much of the functional interrogation remained to be further studied.

**Breast cancer subtype prediction**

While predicting cancers from different anatomic sites may be relative straightforward, predicting cancer subtypes, such as breast cancer, is an ongoing research topic. Breast cancer is divided into four subtypes known as luminal (A&B), HER2 positive and basal (often triple negative breast cancers (TNBC)) breast cancers[32]. In order to accomplish this, we further trained 1D-CNN model with all breast cancer samples from four different subtypes plus the normal breast cancer and set prediction layer with 5 nodes. To further simplify the 1D-CNN, the fully connected layer with 128 nodes was removed. After training, we achieved an average precision of 88.3% (details see **Table 2**). Not that the misclassification is mainly between luminal A & B classes (two inherently similar tumor subtypes), and Her2 class' low classification error is due to the lack of information since it is defined as the gain in DNA copy number (not necessarily reflected in gene expression) and/or over-expression of ERBB2 gene.

## Discussion

There were several critical issues that this paper addressed to improve the accuracy of our prediction and interpretation. Specifically, three CNN architectures were proposed to investigate an appropriate architecture for unstructured gene expressions for predicting cancer types. As is shown in **Figure 2B,** 1D-CNN and 2D-Hybrid-CNN achieved comparable accuracy (95.7%), which improves the result (95.6%) slightly in [9]. Note that 2D-Vanilla-CNN contains only one layer and 32 kernels, whereas the 2D-3Layer-CNN consists of multiple DL modules, a much more complex architecture. In addition to what we summerized in **Table 3** where number of parameters for each model, loss function value after training and testing, and execution time examples, we note several underlying design facts behind each proposed model.

- The *1D-CNN* is significantly simpler than the other models proposed in the literature. It does not require inputs to be arranged in a particular order and it has only one convolutional layer. This much-simplified design induces a significant reduction in the number of hyperparameters (from 26 millions to ~200 thousand) to be estimated during training. This is highly desirable in the DL applications in genomic studies due to the difficulty and the high cost in collecting large genomic data.

- The *2D-Vanilla-CNN* has around one million hyperparameters which are significantly more than those of the 1D-CNN. The model became more difficult to converge when the stride of the kernel was selected to be 1x1. Also, by sliding two separate convolutions kernels over the two orthogonal dimensions, it improved the accuracy due to the ability to capture more global features.

While 2D-Hybrid-CNN may provide a slight advantage in terms of the averaged classification accuracy (**Figure 2B**), it has two times more hyperparameters and thus a higher computation burden compared with the 1D-CNN model. Therefore, we focused on the 1D-CNN model in most of our subsequent analysis.

2D-Vanilla-CNN has similar accuracy comparing to 1D-CNN, but has almost 5x more hyperparameters to train. In order to investigate the robustness of proposed models in presence of noise, both CNN models were tested with data added with different level of noise as explained in Methods section. In **Figure 5**, the 5-fold accuracy of 1D-CNN and 2D-Vanilla-CNN while tested on different ratios of noise are represented. As it is shown, the performance of both models is extremely robust until noise ratio reaches 75% and then it gradually drops. Although both models have almost equal performance results until 75% noise ratio, 1D-CNN outperforms 2D-Vanilla-CNN in noise ratios above 75%. Thus, we can conclude 1D-CNN has more stable performance encountering unwanted noise compared to other models.

We chose to combine tumor samples plus normal samples together to train a DL model with 34 nodes in the prediction layer such that we can eliminate the influence of tissue origin in cancer type prediction. The model not only achieved a good precision in predicting normal tissues (92.5% precision) but made few mistakes in distinguishing cancer types within the same tissue origin; examples include KICH, KIRC, and KIRP, all of them are kidney cancers, where only 2 normal samples are classified into cancer groups (out of 128 normal kidney samples, **Figure 3C**). We will continue our work to investigate this issue by introducing yet another rich source of transcriptomic data from GTEx collection[33]. Furthermore, as suggested by previous studies[13, 15, 34-37], we may incorporate additional genome-wide profiling information, such as DNA mutation, copy number variation, and DNA methylation as additional input matrices to enrich the complexity for model training, and thus to improve the classification accuracy.

Our unique interpretation of the CNN for genomic data has shown its utility when we examined the gene-effect scores. While some of these differences are modest in magnitude, our DL machine had no trouble to classify tumors into their correct subtypes, indicating a simple linear classifier (i.e., expression high vs. low) might not explain the complexity of our CNN. In this sense, our CNN model had the benefit of capturing high-order interactions among these genes to make accurate predictions.

## Conclusions

Taken together, we have presented three unique CNN architectures that take high dimension gene expression inputs and perform cancer type prediction while considering their tissue of origin. Our model achieved an equivalent 95.7% prediction accuracy comparing to earlier published studies, however with a drastically simplified CNN construction and with a significant reduction from tissue of origin. This allows us to perform a normal interpretation of our CNN model to elucidate cancer markers for each cancer type, with hope in future refinement that will lead to markers for earlier cancer detection.

## List of Abbreviations

ACC, adrenocortical cancer; BLCA, bladder urothelial carcinoma; BRCA, breast invasive carcinoma; CESC, cervical and endocervical cancer; CHOL, cholangiocarcinoma; COAD, colon adenocarcinoma; CNN, convolutional neural network; DL, deep learning; DLBC, diffuse large B-cell lymphoma; ESCA, esophageal carcinoma; GBM, glioblastoma multiforme; HNSC, head and neck squamous cell carcinoma; KICH, kidney chromophobe; KIRC, kidney clear cell carcinoma; KIRP, kidney papillary cell carcinoma; LAML, acute myeloid leukemia; LGG, lower grade glioma; LIHC, liver hepatocellular carcinoma; LUAD, lung adenocarcinoma; LUSC, lung squamous cell carcinoma; MESO, mesothelioma; OV, ovarian serous cystadenocarcinoma; *P*, *P*-value; PAAD, Pancreatic Adenocarcinoma; PCPG, pheochromocytoma and paraganglioma; PRAD, prostate adenocarcinoma; READ, rectum adenocarcinoma; SARC, sarcoma; SKCM, skin cutaneous melanoma; STAD, stomach adenocarcinoma; TCGA, The Cancer Genome Atlas; TGCT, testicular germ cell tumor; THCA, thyroid carcinoma; THYM, thymoma; UCEC, uterine corpus endometrioid carcinoma; UCS, uterine carcinosarcoma; UVM, uveal melanoma;

# Declarations

**Ethics approval and consent to participate**

Not applicable.

**Consent for publication**

Not applicable.

**Availability of data and material**

The dataset supporting the conclusions of this article is included within the article.

**Competing interests**

The authors declare that they have no competing interests.

**Funding**

This research and this article's publication costs were supported partially by the NCI Cancer Center Shared Resources (NIH-NCI P30CA54174 to YC), NIH (CTSA 1UL1RR025767-01 to YC, and R01GM113245 to YH), CPRIT (RP160732 to YC and MM), and San Antonio Life Science Institute (SALSI Innovation Challenge Award 2016 to YH and YC and SALSI Postdoctoral Research Fellowship to YCC). The funding sources had no role in the design of the study and collection, analysis, and interpretation of data and in writing the manuscript.

**Authors' contributions**

All of the authors conceived the study. MM and YCC designed the model and performed data analysis. All authors interpreted the data and wrote the manuscript. All of the authors have read and approved the final manuscript.

**Acknowledgements**

The authors greatly appreciate the intensive discussion and constructive suggestions with all members from Machine Learning Interesting Group organized by Drs. Huang and Chen.

# Tables

**Table 1A**. Different hyperparameter settings for 1D-CNN model based on the trained and tested statistical measures. The final selected parameters are highlighted.

| Hyperparameters | | | Loss | | | |
|---|---|---|---|---|---|---|
| dense layer size | filter | kernel | mean train_score | stdev train_score | mean test_score | stdev test_score |
| 64 | (1, 50) | 8 | 0.069 | 0.031 | 0.167 | 0.023 |
| 64 | (1, 50) | 16 | 0.037 | 0.013 | 0.140 | 0.007 |
| 64 | (1, 50) | 32 | 0.023 | 0.003 | 0.132 | 0.006 |
| 64 | (1, 50) | 64 | 0.013 | 0.002 | 0.128 | 0.006 |
| 128 | (1, 50) | 8 | 0.032 | 0.008 | 0.147 | 0.006 |
| 128 | (1, 50) | 16 | 0.027 | 0.014 | 0.138 | 0.014 |
| 128 | (1, 50) | 32 | 0.011 | 0.003 | 0.121 | 0.009 |
| 128 | (1, 50) | 64 | 0.004 | 0.001 | 0.126 | 0.012 |
| 512 | (1, 50) | 8 | 0.009 | 0.000 | 0.138 | 0.008 |
| 512 | (1, 50) | 16 | 0.006 | 0.001 | 0.127 | 0.003 |
| 512 | (1, 50) | 32 | 0.124 | 0.179 | 0.265 | 0.160 |
| 512 | (1, 50) | 64 | 0.003 | 0.002 | 0.125 | 0.008 |
| 64 | (1, 71) | 8 | 0.072 | 0.009 | 0.177 | 0.009 |
| 64 | (1, 71) | 16 | 0.044 | 0.009 | 0.149 | 0.006 |
| 64 | (1, 71) | 32 | 0.036 | 0.011 | 0.135 | 0.009 |
| 64 | (1, 71) | 64 | 0.016 | 0.004 | 0.124 | 0.012 |
| 128 | (1, 71) | 8 | 0.046 | 0.007 | 0.154 | 0.015 |
| 128 | (1, 71) | 16 | 0.027 | 0.006 | 0.135 | 0.015 |
| **128** | **(1, 71)** | **32** | **0.014** | **0.002** | **0.129** | **0.016** |
| 128 | (1, 71) | 64 | 0.008 | 0.001 | 0.119 | 0.003 |
| 512 | (1, 71) | 8 | 0.023 | 0.018 | 0.152 | 0.023 |
| 512 | (1, 71) | 16 | 0.009 | 0.008 | 0.132 | 0.017 |
| 512 | (1, 71) | 32 | 0.004 | 0.002 | 0.123 | 0.008 |
| 512 | (1, 71) | 64 | 0.011 | 0.016 | 0.134 | 0.015 |
| 64 | (1, 100) | 8 | 0.088 | 0.010 | 0.172 | 0.015 |
| 64 | (1, 100) | 16 | 0.066 | 0.014 | 0.162 | 0.009 |
| 64 | (1, 100) | 32 | 0.037 | 0.007 | 0.132 | 0.009 |
| 64 | (1, 100) | 64 | 0.024 | 0.009 | 0.128 | 0.013 |
| 128 | (1, 100) | 8 | 0.058 | 0.001 | 0.164 | 0.009 |
| 128 | (1, 100) | 16 | 0.031 | 0.008 | 0.144 | 0.014 |
| **128** | **(1, 100)** | **32** | **0.019** | **0.004** | **0.128** | **0.008** |
| 128 | (1, 100) | 64 | 0.016 | 0.010 | 0.137 | 0.027 |
| 512 | (1, 100) | 8 | 0.031 | 0.013 | 0.155 | 0.014 |
| 512 | (1, 100) | 16 | 0.009 | 0.001 | 0.135 | 0.009 |

**Table 1B.** Different hyperparameter settings for 2D-Vanilla-CNN model based on the trained and tested statistical measures. The final selected parameters are highlighted.

| Hyperparameters | | | | Loss | | | |
|---|---|---|---|---|---|---|---|
| dense layer size | filter | kernel | stride | mean train_score | stdev train_score | mean test_score | stdev test_score |
| 128 | 32 | (7, 7) | (1, 1) | 20.999 | 18.228 | 21.281 | 14.904 |
| 128 | 32 | (7, 7) | (2, 2) | 0.005 | 0.002 | 0.192 | 0.022 |
| 128 | 32 | (10, 10) | (1, 1) | 21.398 | 18.582 | 21.771 | 15.298 |
| **128** | **32** | **(10, 10)** | **(2, 2)** | **0.009** | **0.003** | **0.187** | **0.008** |
| 128 | 32 | (20, 20) | (1, 1) | 0.027 | 0.004 | 0.202 | 0.029 |
| 128 | 32 | (20, 20) | (2, 2) | 0.043 | 0.011 | 0.206 | 0.009 |
| 128 | 64 | (7, 7) | (1, 1) | 10.213 | 17.688 | 10.566 | 14.618 |
| 128 | 64 | (7, 7) | (2, 2) | 0.004 | 0.001 | 0.187 | 0.018 |
| 128 | 64 | (10, 10) | (1, 1) | 31.430 | 1.149 | 31.675 | 1.019 |
| 128 | 64 | (10, 10) | (2, 2) | 0.012 | 0.006 | 0.177 | 0.014 |
| 128 | 64 | (20, 20) | (1, 1) | 12.020 | 18.052 | 12.149 | 14.818 |
| 128 | 64 | (20, 20) | (2, 2) | 0.055 | 0.016 | 0.204 | 0.020 |
| 512 | 32 | (7, 7) | (1, 1) | 21.245 | 18.419 | 21.175 | 14.815 |
| 512 | 32 | (7, 7) | (2, 2) | 10.944 | 18.953 | 11.022 | 15.306 |
| 512 | 32 | (10, 10) | (1, 1) | 10.964 | 18.987 | 11.148 | 15.482 |
| 512 | 32 | (10, 10) | (2, 2) | 0.003 | 0.001 | 0.213 | 0.025 |
| 512 | 32 | (20, 20) | (1, 1) | 10.988 | 19.002 | 11.132 | 15.436 |
| 512 | 32 | (20, 20) | (2, 2) | 1.110 | 1.849 | 1.271 | 1.397 |
| 512 | 64 | (7, 7) | (1, 1) | 31.430 | 1.149 | 31.675 | 1.019 |
| 512 | 64 | (7, 7) | (2, 2) | 10.213 | 17.688 | 10.560 | 14.622 |
| 512 | 64 | (10, 10) | (1, 1) | 31.497 | 1.211 | 31.648 | 1.087 |
| 512 | 64 | (10, 10) | (2, 2) | 20.628 | 17.858 | 20.481 | 14.363 |
| 512 | 64 | (20, 20) | (1, 1) | 11.299 | 16.825 | 11.562 | 13.969 |
| 512 | 64 | (20, 20) | (2, 2) | 12.020 | 18.046 | 12.152 | 14.776 |

**Table 2. Breast cancer subtype classification using 1D-CNN model**

| Class name | Precision | Recall | F1-score | Support |
|---|---|---|---|---|
| **Basal** | 0.973 | 0.980 | 0.976 | 147 |
| **Her2** | 0.829 | 0.853 | 0.841 | 68 |
| **Luminal A** | 0.894 | 0.927 | 0.910 | 437 |
| **Luminal B** | 0.810 | 0.780 | 0.795 | 186 |
| **Normal** | 0.857 | 0.462 | 0.600 | 26 |
| **Avg/Total** | 0.883 | 0.884 | 0.882 | 864 |

**Table 3. Hyperparameters and training times of 4 CNN models**

| DL model[1] | Number of parameters | Training Loss | Training Accuracy | Testing Loss | Testing Accuracy[2] | Time[3] (seconds) |
|---|---|---|---|---|---|---|
| 1D-CNN | 211,489 | 0.01 | 0.9971 | 0.1769 | 0.9567 | 80.3 |
| 2D-Vanilla-CNN | 1,420,737 | 0.007 | 0.9981 | 0.1778 | 0.9557 | 94 |
| 2D-Hybrid-CNN | 362,177 | 0.0149 | 0.996 | 0.1586 | 0.9582 | 80.8 |
| 2D-3Layer-CNN | 26,211,233 | 0.5149 | 0.9654 | 0.6875 | 0.9184 | 214.6 |
| 2D-3Layer-CNN (with patience = 10) | | 0.1976 | 0.9869 | 0.3914 | 0.9419 | 379.17 |

1. Early stopping is used for all models (all with patience = 4, except for the last model.)
2. This is the result of 5-cv
3. All DL model training were performed in a Linux server with Xeon 8176 CPU @2.1GHz, with 4x28 cores.

# Figures

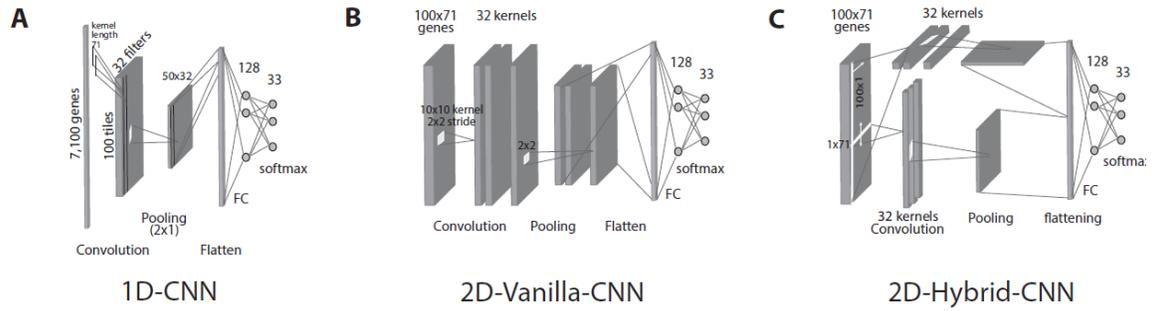

**Figure 1 - Illustration of three CNN models**

(A) 1D-CNN with input as a vector format with 7100 genes. (B) 2D-Vanilla-CNN, with an input reformatted as a 100x71 matrix, and one convolution layer. (C) 2D-Hybrid-CNN, similar input as in (B) but with two parallel convolution layers, vertical and horizontal, as in (A). (D) The performance of 1D-CNN and 2D-Vanilla-CNN when they are tested on different noise ratios added to the inputs.

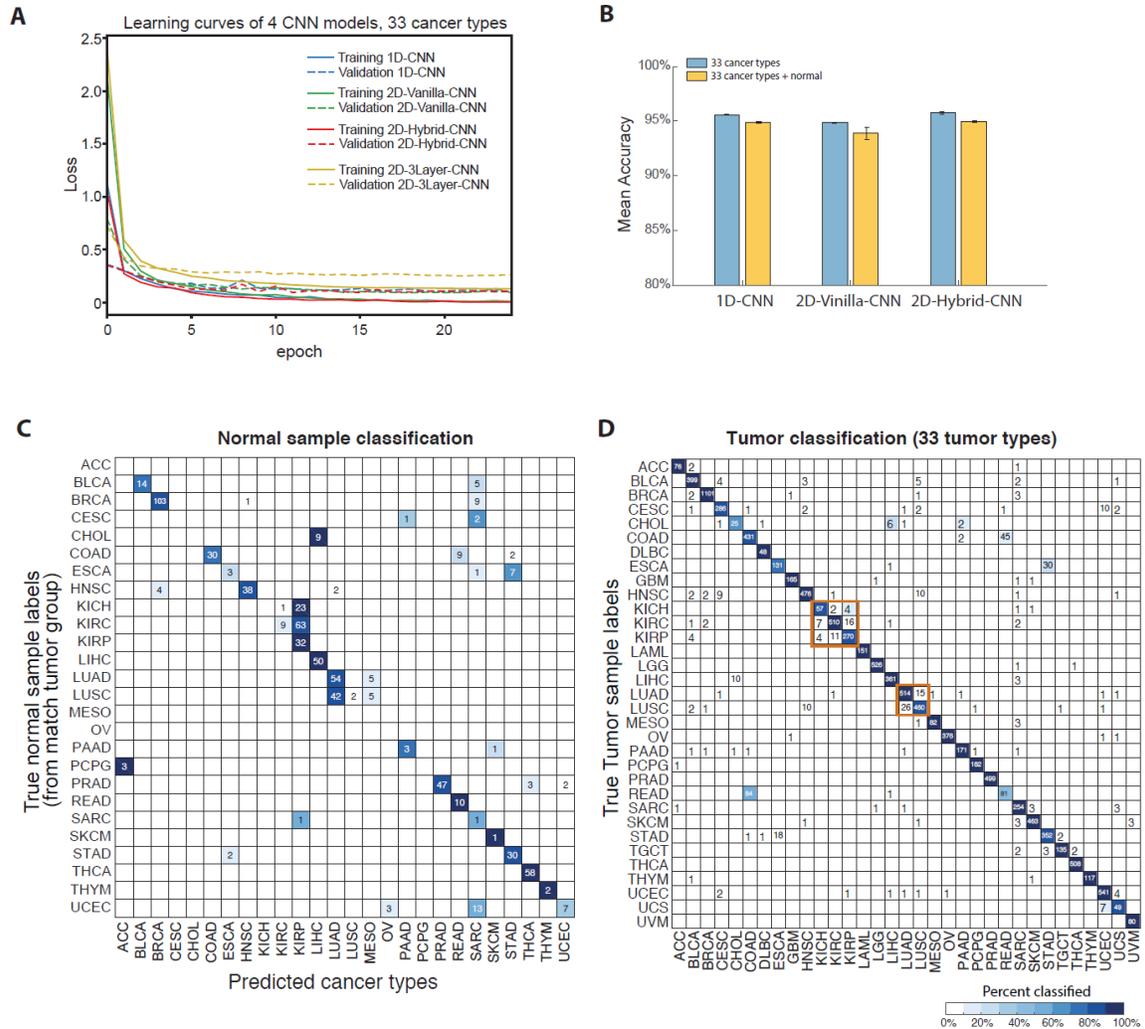

**Figure 2 - Cancer type prediction performance of three CNN models trained with tumor samples only.**

(A) Learning curves for all three CNN models. (B) Micro-averaged accuracy of three CNN models when trained with only tumor samples (light blue) from 33 tumor types, and with tumors and normal samples together (light brown). (C) Confusion matrix of normal samples prediction from 1D-CNN model trained with 33 tumor types only.

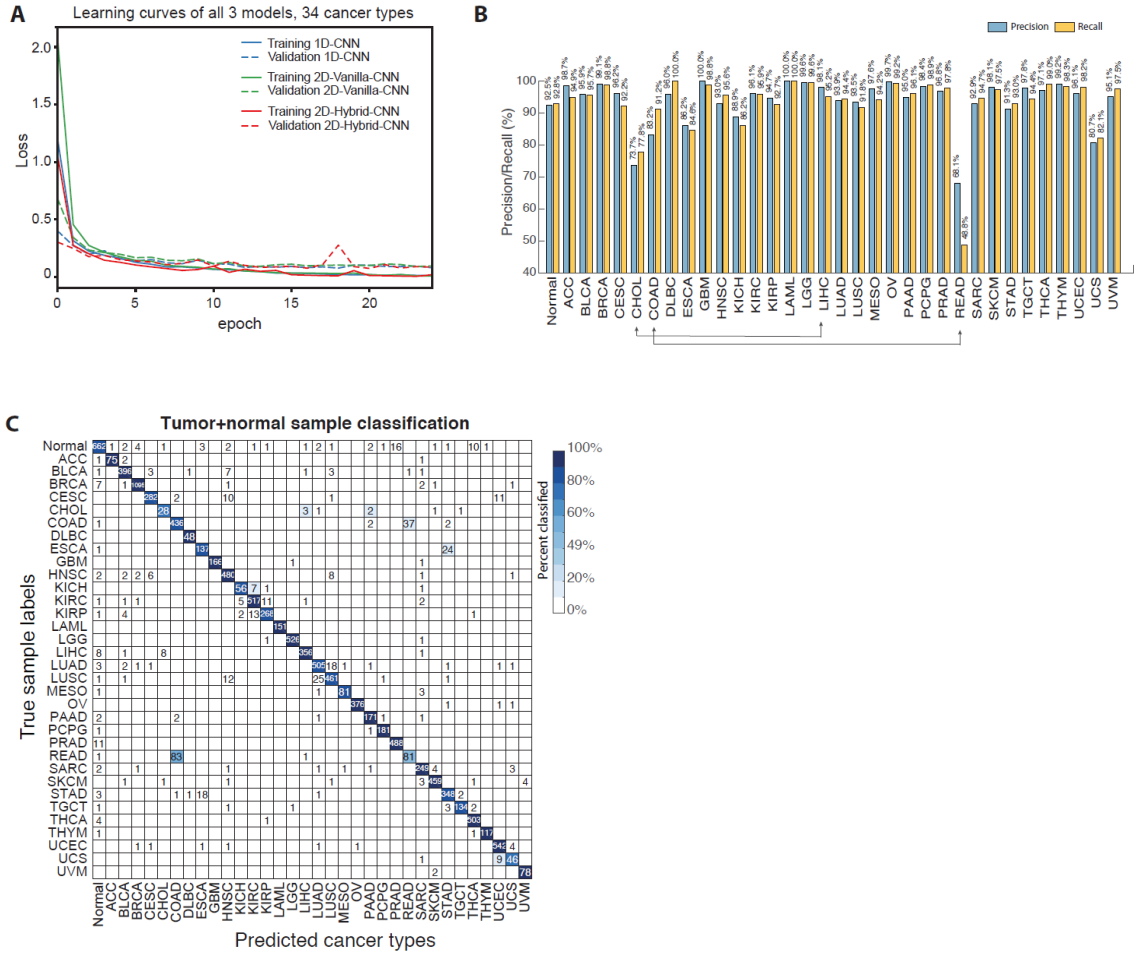

**Figure 3 - Cancer type prediction performance of three CNN models trained with combined tumor and normal samples.**

(A) Learning curves for all three CNN models trained with combined tumor and normal samples. (B) Precision (light blue) and recall (light brown) of 1D-CNN model when trained with combined tumor normal samples. (C) Confusion matrix of all sample prediction from 1D-CNN model trained with 33 tumor types + normal.

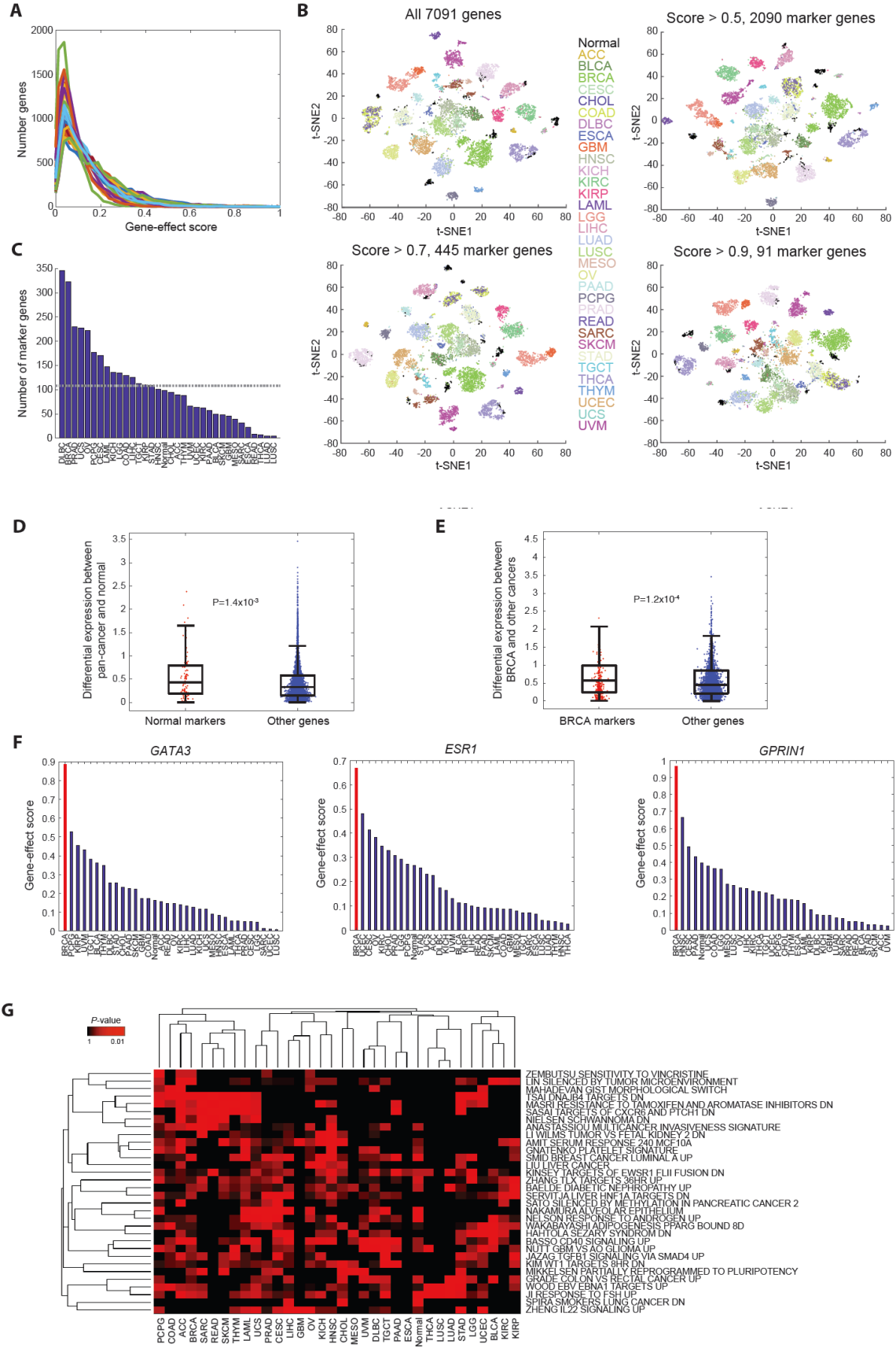

**Figure 4 - Interpretation of the 1D-CNN model**

(A) Distributions of gene-effect scores for individual cancer and normal classes. (B) t-SNE plots of pan-cancer and normal samples by expression of marker genes identified using different thresholds. (C) Marker genes identified in each class with a criterion of gene-effect score > 0.5. (D-E) Differential expression of marker genes and other genes between sample classes. Here differential expression is presented by an absolute difference between a class (normal or BRCA) and all other samples in $\log_2(FPKM+1)$. (F) Pan-classes gene-effect scores of three marker genes of BRCA. (G) Functions associated with marker genes identified in each class.

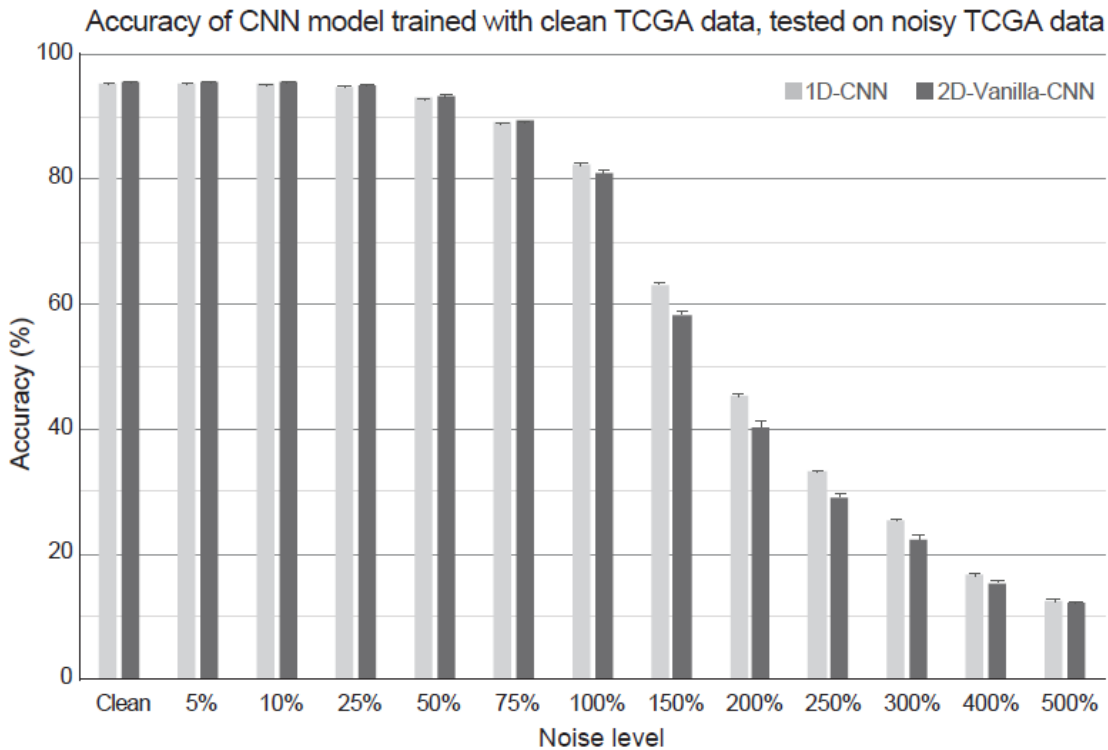

**Figure 5 - CNN models testing on noisy data.**

Classification accuracy on TCGA data with different additive Gaussian noise added. Both classifiers were trained with original TCGA data, but tested on TCGA data + Gaussian noise.